\documentclass[modern,twocolumn]{aastex7}
\usepackage{ulem}

\accepted{January 9, 2026}
\usepackage{graphicx}
\usepackage{amsmath}	
\usepackage{bm}
\usepackage{subcaption}
\usepackage{natbib}
\usepackage{hyperref}
\usepackage{ulem}
\usepackage{booktabs}
\usepackage[a4paper, total={6in, 9in}]{geometry}
\hypersetup{
    colorlinks=true,
    linkcolor=blue,
    filecolor=magenta,      
    urlcolor=blue,
    citecolor=blue,
    linkcolor=blue,
    pdftitle={Chitan OJ Disk Paper},
    }

\begin{document}

\title{Long Term (250 years) Hydrodynamical Simulation of the Supermassive Black Hole Binary OJ287}

\author[orcid=0000-0001-7163-6346]{Ariel Chitan}
\affiliation{Department of Physics and Astronomy, Western University}
\email[show]{achitan2@uwo.ca}
\affiliation{Institute for Earth and Space Exploration, Western University}

\author[orcid=0000-0001-6217-8101]{Sarah C. Gallagher}
\affiliation{Department of Physics and Astronomy, Western University}
\affiliation{Institute for Earth and Space Exploration, Western University}

\email{sgalla4@uwo.ca}

\author[orcid=0000-0003-0428-2140]{Shahram Abbassi}
\affiliation{Department of Physics and Astronomy, Western University}

\email{sabbassi@uwo.ca}


\begin{abstract}
With upcoming facilities capable of detecting photometric and gravitational wave signals from supermassive black hole (SMBH) binaries, studying their long-term accretion-driven variability is timely. OJ287 is a bright, nearby ($z=0.3$), and well-studied candidate for a SMBH binary. As such, it is an excellent case study for how binary dynamics could influence observed active galactic nucleus (AGN) photometric variability. We present 3D hydrodynamic simulations of OJ287, using the code \scriptsize PHANTOM\normalsize. We simulate two mass ratios: (i) M$_1$ $=$  1.835$\times$10$^{10}$ M$_\odot$ with M$_2$ $=$ 1.4$\times$10$^{8}$ M$_\odot$, (ii) M$_1\approx$ M$_2$ ($\sim10^{8}$ M$_\odot$) along and (iii) control of a single SMBH and accretion disc. We find that the simulation with masses 1.835$\times$10$^{10}$ M$_\odot$ and 1.4$\times$10$^{8}$ M$_\odot$ evolves consistently with the most currently accepted model of OJ287 as a precessing SMBH binary. The secondary's impacts with the disc result in the formation of spiral density waves and a corresponding $\sim$10--20\% increases in the mass accretion rate of the primary SMBH. The impact timings and the mass accretion rate spikes show quasi-periodic variability as a result of the precession of the secondary's orbit with intervals between impacts ranging from $\sim$ 1 year to $\sim$ 10 years. In the near-equal mass case, the disc of the primary becomes tidally disrupted after $\sim$ 2 years. Consequently, the near-equal mass system  with a period of 12 years is not a viable candidate for OJ287.  This modeling provides insights into the potential signatures of SMBH binaries by both gravitational wave observatories and the Rubin Legacy Survey of Space and Time.

\end{abstract}



\section{Introduction}
\label{sect:intro}
SMBHs are fundamental components of galaxies found in galactic cores (e.g., \citealt{kormendy1995} and references within). With masses from 10$^6$--10$^{11}$ M$_\odot$ 
\citep{greene2007,melo-carneiro2025},
how these SMBHs grew is a key question in the cosmological landscape. SMBHs once formed are thought to grow rapidly in the early universe via (i) a series of mergers with other SMBHs due to the hierarchical merging of galaxies and their halos and (ii) the accretion of matter at up to super-Eddington rates \citep[e.g.,][]{Marta2010}.

Though mergers of SMBHs have not yet been observed, candidate SMBH binaries with small separations have been detected from the periodic variability of their photometric lightcurves. One example, PG 1302--102, has a period of $\sim$5 years and SMBHs with a total estimated virial mass of 10$^{8.5}$ M$_\odot$ (found using single epoch virial black hole estimation) with a separation of $\sim$0.01 parsecs \citep{graham2015a}. 
 SDSS J0159+0105 is another such candidate, with a period of $\sim$4 years and total proposed virial mass of 10$^{8.5}$ M$_\odot$, identified by its periodic lightcurve \citep{zheng2016}. 
 
 However, even with strong evidence of binary signatures, most binary candidates remain unconfirmed. Candidates like PG 1302--102 for example, have been scrutinized in detail. \citet{vaughan2016} conducted a Bayesian analysis of the lightcurve of PG 1302--102 and showed that a simple, stochastic model fit better than a sinusoidal one, the signature that was used initially to categorize the object as a binary candidate. This highlights the difficulty in identifying which binary candidates are actually binaries or just false positives particularly over limited timeframes. 
 
More recently, \citet{huijse2025} identified a sample of 181 candidate SMBH binaries from the Gaia DR3 \citep{gaia2023} by looking for periodicity in the lightcurves of sources in a time range between 100 and 1000 days. However, \citet{elbadry2025}, used data from the Zwicky Transient Facility \citep[ZTF,][]{bellm2019} to extend this period from 1000 days to 4000 days for 116 of the candidates that were available in ZTF and found that in all of the 116 candidates the Gaia periodic model failed to further predict periodic variability - only stochastic. They state that finding true periodic signatures coming from SMBH binaries is very rare.


Large samples of candidate SMBH binaries are still worth producing as they can be potential sources of gravitational wave emission that could be detectable with current and upcoming observatories.  To that end, \citet{graham2015b} have identified 111 candidate SMBH binaries from the Catalina Real-time Transient Survey. An additional 33 from the Palomar Transient Factory were found by \citet{charisi2016}. Furthermore, \citet{sesana2018} have shown that using the Catalina and Palomar SMBH binary candidates to estimate the SMBH merger rates results in inconsistent gravitational wave background (exceeds the observational constraints by an order of magnitude) with that measured from pulsar timing arrays, suggesting an overestimation of the number of SMBH binaries from the candidate binary catalogs. Additionally, \citet{sydnor2022} is presenting a plan for a project to construct a comprehensive a publicly available, searchable catalog containing candidate SMBH binaries, the Black holes Orbiting Black holes catalog (BOBcat). Further detailed analyses of these large binary candidate catalogs will be necessary.

One well-studied SMBH binary candidate is the blazar OJ287 \citep{sillanpaa1988}, with recent detection of jet activity from the secondary strengthening its binary status \citep{valtonen2024,valtonen2025}. OJ287 is one of the most famous candidates for a SMBH binary due to its distinctive optical variability that features optical flares dating back to the early 1900's (see \citealt{hudec1913} for a comprehensive history). Because this object has been well documented for over 100 years, it is our best candidate for a SMBH binary and presents a unique opportunity to study such a system in great detail.

\begin{figure*}
     \centering
     \includegraphics[width=0.75\linewidth]{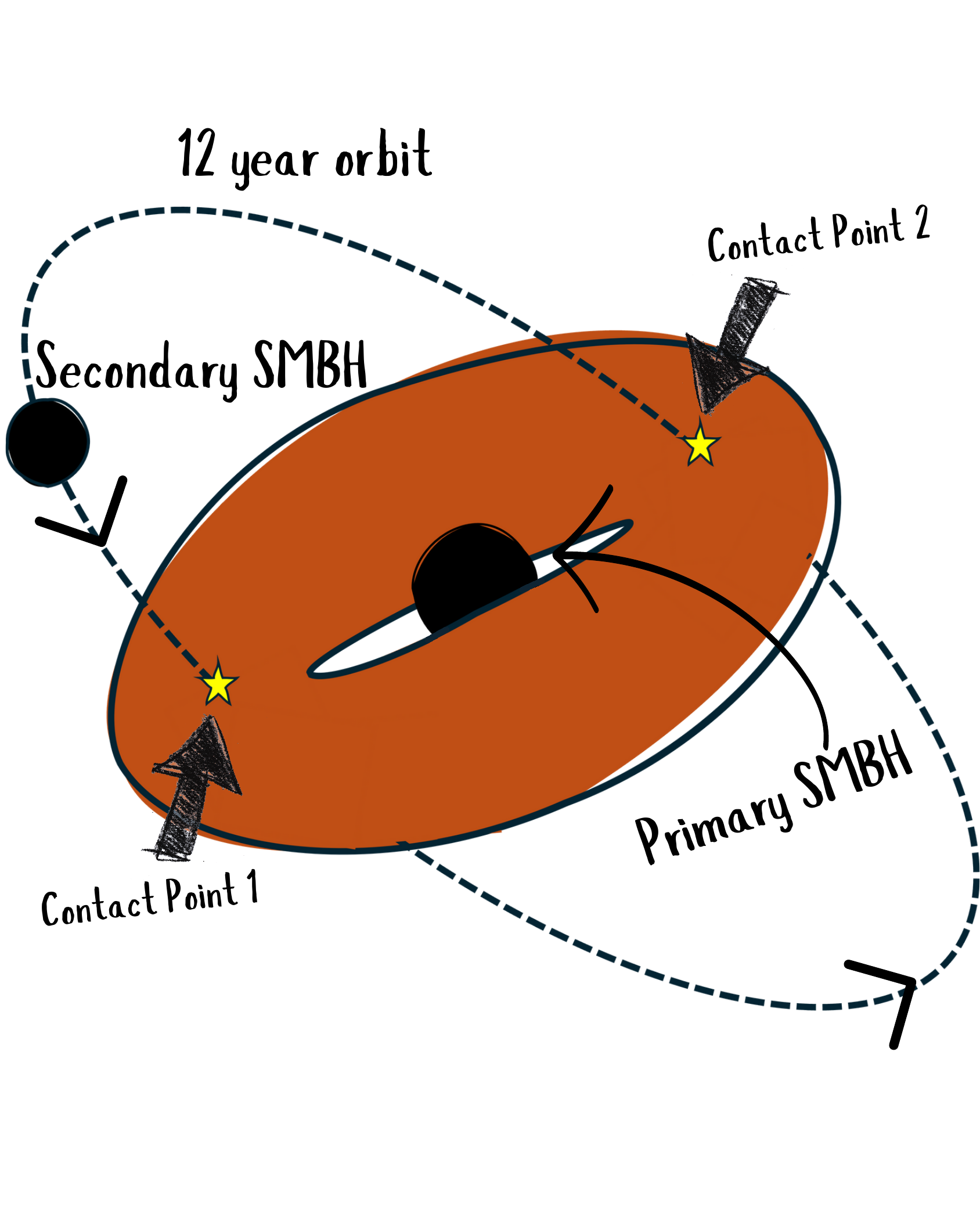}
     \caption{A diagram of the geometry of the OJ287 binary black hole system. The secondary SMBH orbits around the primary with a 12 year period. As the secondary passes through the disc of the primary along its orbit (Contact Points), it produces detectable flares.}
     \label{fig:ojschematic}
 \end{figure*}

This quasi-periodic BL Lacertae object at a redshift of $z= 0.306$ produces detectable flares in the optical with a period of $\sim$ 12 years. The quasi-periodicity of OJ287 is observed at multiple wavelengths \citep{prince2021}. The flares are proposed to originate from the secondary's impact with the primary's accretion disc \citep{sillanpaa1988}. A schematic of the proposed geometry of the OJ287 system is shown in Figure \ref{fig:ojschematic}.

\begin{figure}
    \centering
    \includegraphics[width=\linewidth]{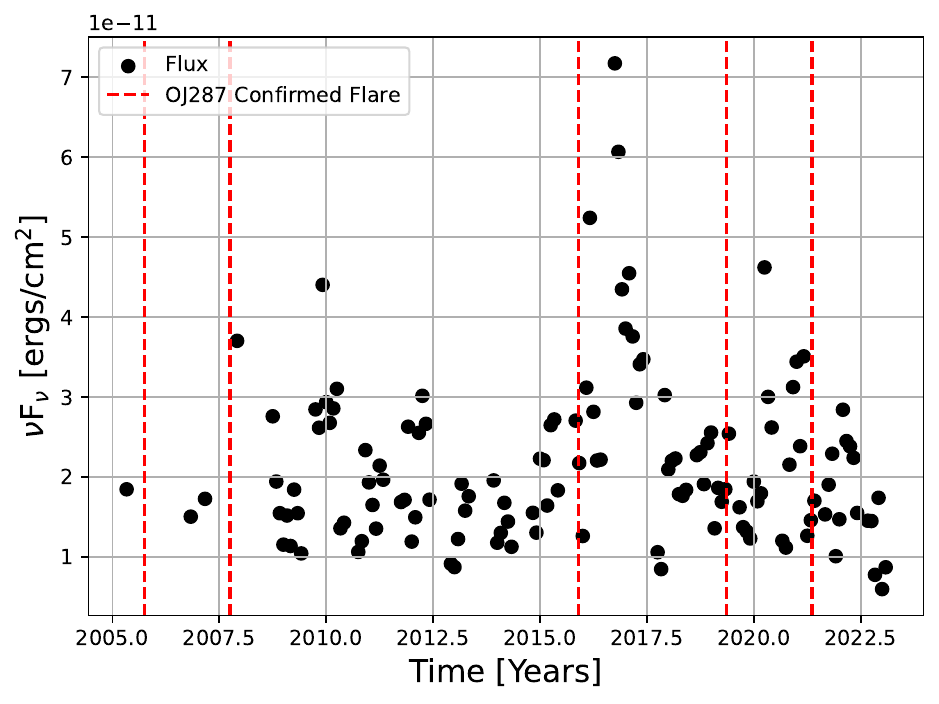}
    \caption{The observed optical flux for the OJ287 system from 2005--2022. The wavelength range for this data was $461.5$nm $\leq \lambda \leq 428.6$nm, taken from the MMDC \citep{sahakyan2024} catalog. It also contains data from \citet{bonning2012}. The vertical red lines show when there was a confirmed observed flare from the system. These are 2005.76 \citep{valtonen2008}, 2007.69 \citep{valtonen2008nature}, 2015.87 \citep{valtonen2016}, 2019.36 \citep{laine2020} and 2021.9 \citep{valtonen2024}. Some of these indicate the beginning of the flares.}
    \label{fig:observedflux}
\end{figure}

For reference, the observed optical lightcurve for OJ287 from 2005--2022.5 is shown in Figure \ref{fig:observedflux} (data from the Markarian Multiwavelength Data Center\footnote{https://mmdc.am/} \citealt{sahakyan2024}).  The strong flaring activity (a factor of $\sim3$) in 2016 is characteristic of the behaviour of OJ287 over the past 100+ years \citep{sillanpaa1988,hudec1913}. Recent (with ZTF) and upcoming (with Rubin) high-cadence photometric observations offer the promise to provide more empirical constraints on the system.

\citet{lehto1996} first proposed OJ287 to be a precessing SMBH binary. They found that for the flares to accurately match up to a binary model, it was necessary that the secondary's orbit be precessing relativistically. Relativistic periapsis precession would affect the timing of the optical flares. The most recent estimates for the masses of the black holes are $\sim10^{10}$~M$_\odot$ for the primary and $\sim10^{8}$~M$_\odot$ for the secondary \citep{valtonen2023}. In this scenario, the secondary SMBH punches through the accretion disc of the primary, shock heating the gas and causing it to expand out of the accretion disc \citep{ivanov1998}, giving rise to bremsstrahlung radiation and the observed flares \citep{gopal-krishna2024}. 

Relativistic precession of the orbit predicts that at some point, the orbit of the secondary would lie close to parallel to the plane of the accretion disc \citep{valtonen2023}. During these approaches, the timing of the impacts are subject to greater variability as the disc itself becomes warped due to the proximity of the secondary throughout its orbit. 

Recently, the mass of the OJ287 system has been debated. Because OJ287 was not observable from ground-based telescopes in late 2022, the detection (or not) of a predicted flare did not occur \citep{valtonen2024}. \citet{valtonen2024b} showed that this particular flare would need greater precision to pinpoint in time \citep{valtonen2021} because of the alignment of the secondary's orbit parallel to the plane of the primary's accretion disc, a natural consequence of a precessing orbit. \citet{valtonen2021} describe how the disc becomes warped and the impact times need to be more precisely calculated when the secondary's orbit lies along the disc.  According to \citet{valtonen2024b}, the orbital alignment coupled with OJ287 not being visible during certain flaring periods, there was no missing flare of OJ287 in 2022 but rather a missed detection of it. \citet{komossa2023} propose instead that a missed 2022 flare indicates that the mass of the primary SMBH could instead be 10$^8$ M$_\odot$, two orders of magnitude less than predicted by the precessing model. 

Though the most widely accepted model for OJ287 is a SMBH binary, alternate theories have been proposed by \citet{britzen2018} and \citet{butuzova2020} that attribute the quasi-periodicity of the system to the jet of the primary. \citet{britzen2018}, in particular, show how the precession and nutation of the jet, with a modeled timescale of about 24 years, can lead to both radio and optical flares without the need for a secondary SMBH using 120 Very Long Baseline Array observations at 15Ghz. \citet{britzen2018} discuss how the precession of the jet can be due to either wobbling of the disc (Lense-Thirring precession) or binary perturbations from a secondary SMBH. However, they state that a binary nature is a more reasonable explanation. 

Regardless of these questions, OJ287 remains one of the most studied candidate SMBH binary systems and provides a well-documented template for future SMBH binary detections using upcoming facilities. The upcoming Legacy Survey of Space and Time (LSST) by the Rubin Observatory is expected to detect $\gtrsim10^6$ SMBH binaries candidates from their photometric variability characteristics \citep{xin2021}.
An in-depth investigation into a well-characterized system such as OJ287 can serve as a sample template for these surveys as well as offer further insight into the properties of the system.
OJ287 is also an ideal candidate to illustrate the mechanisms of merger and accretion for the growth of SMBHs: it will soon (astronomically speaking) merge and is also currently an AGN. 

In this paper, we present results from long-term hydrodynamical simulations of OJ287. The general relativistic magnetohydrodynamics simulations of OJ287 from \citet{ressler2025} model the short-time impacts of the secondary passing through the disc. They used the code Athena++ \citep{white2016} and looked at three different mass ratios of the binary with q=0.025, 0.05 and 0.1.  \citet{ressler2025} specifically investigate the impact events and the physical nature of the induced flares. They also find that twice during their simulation, the impact causes the secondary to temporarily form its own jet. In contrast, we conduct hydrodynamic simulations of OJ287 over 250 years, modeling a single accretion disc surrounding the primary. Both SMBHs and the primary accretion disc were simulated using the 3D smoothed particle hydrodynamics code, \scriptsize PHANTOM \normalsize \citep{price2018}. We investigate how this system evolves using two mass scenarios, one in which the secondary is $\sim1\%$ the mass of the $\sim10^{10}$~M$_{\odot}$ primary, and the second with two $\sim10^{8}$~M$_{\odot}$ black holes. 

This paper is separated into the following sections: in Section \ref{sect:setup} we develop the initial conditions and describe each of the simulations in greater detail, in Section \ref{sect:results} we present the results of the simulations and in Section \ref{sect:discussion}, we explore some of the astrophysical implications of our findings.

\section{Simulation Setup}\label{sect:setup}

\scriptsize PHANTOM \normalsize       \citep{price2018} is a widely used, publicly available, smoothed particle hydrodynamics code based on the theoretical framework of \citet{lucy1977} and \citet{ginghold1977}. It uses a Lagrangian approach to fluid dynamics. Particles are moved along the local fluid velocity field during the simulation.  The accretion disc is discretized into individual particles of equal mass which comprise the initial disc with a mass set by the user.  Here we use it to numerically model the OJ287 system. 

In \scriptsize PHANTOM, \normalsize we use the setup of a sink particle (the secondary) in orbit around another sink particle (the primary) embedded in a disc. We use N$= 4 \times 10^{5}$ particles in each simulation. This setup includes a general relativistic prescription \citep{liptai2019}, which numerically models the apsidal precession necessary to simulate the precessing SMBH binary model.

We follow \citet{lehto1996} for the parameters of the OJ287 system (see Table \ref{tab:parameters}) based on their best fit to the observed pattern of optical flaring. They have found that for this model the orbital period of the secondary is $\sim$12 years (corresponding to a separation of 2.25$\times$10$^4$AU with primary mass 1.835$\times$10$^{10}$ M$_\odot$ and secondary mass 1.4$\times$10$^{8}$ M$_\odot$) and the eccentricity is 0.7. The spin of the primary SMBH, 0.313, is taken from \citet{valtonen2016}, while the viscosity parameter, 0.26, is taken from \citet{valtonen2019}.

\begin{deluxetable}{cc}

\tablehead{
\colhead{Orbital parameter} & \colhead{Value}
}
\tablecaption{Orbital Parameters of OJ287\label{tab:parameters}}

\startdata
Period & 12 years \\
Eccentricity & 0.7 \\
Spin of M$_1$ & 0.313 \\
Disc Viscosity & 0.26 \\
$p_{index}$&1.5\\
$q_{index}$&0.75\\
$\gamma$&1.33
\enddata
\tablecomments{Disc parameters for period, eccentricity, spin and disc viscosity taken from \citet{lehto1996}. $p_{index}$ is the power law index of surface density profile, $q_{index}$ is the power law index of sound speed index and $\gamma$ is the adiabatic gamma.  These values are the same in each simulation.}
\end{deluxetable}

For our discs, we use a thin-disc approximation following \citet{shakura1973}. While this is not the best prescription for a black hole accretion disc \citep{gu2007}, it allows for an analytic description of the disc. As the main focus of our study is to compare the possible impacts of the two different mass scenarios proposed for OJ287, a more complicated disc model is not warranted. The thin disc model was used to approximate the mass of the accretion disc of the primary SMBH in OJ287 

\begin{align}
    M_{\text{disc}} \sim\ & 1.0 \times 10^{-20} \alpha^{4/5} \dot{M}_{17}^{7/10} M_1^{1/4} \nonumber \\
    & \times \left(R^{5/4} - R_{\text{in}}^{5/4}\right) \rm{M_{\odot}}
    \label{eq:discmass}
\end{align}

Here, $\alpha$ is the disc viscosity, $\dot M_{17}$ is $\dot M/10^{17}\mathrm{g\,s^{-1}}$ the mass accretion rate, $M_1$ the primary mass, $R$ is the outer radius of the disc and $R_{\rm in}$, the inner radius of the disc

Further to this, \citet{morgan2010} measured the disc size of quasars via micro-lensing and found that the scaling relation between black hole mass and radius of the disc at a rest frame wavelength of 2500 \r{A} correlated with standard thin disc theory. We use their relation for $R_{2500}$ to initialise the outer radii of our discs :

\begin{equation}
    \log\left(\frac{R_{2500}}{\rm{cm}}\right)=15.78+0.8\log\left(\frac{M_{\bullet}}{10^9 \rm{M_{\odot}}}\right)\label{eq:discradius}
\end{equation}

For the inner radius, we use the innermost stable circular orbit (isco) for a rotating black hole \citep{bardeen1972}:  

\begin{equation}
    R_{\rm{isco}}=\frac{9GM_{\bullet}}{c^2}\label{eq:iscobhwspin}
\end{equation}

\begin{deluxetable*}{cccc}
\tablecaption{Summary of Simulations\label{tab:mass_setups}}

\tablehead{
\colhead{Parameter} & \colhead{Simulation A} & \colhead{Simulation B} & \colhead{Control}
}
\startdata
M$_1$ [M$_\odot$] & 1.84 $\times$ 10$^{10}$ & 1.84 $\times$ 10$^{8}$ & 1.84 $\times$ 10$^{10}$ \\
M$_2$ [M$_\odot$] & 1.40 $\times$ 10$^{8}$ & 1.40 $\times$ 10$^{8}$ & - \\
q & 7.63 $\times$ 10$^{-3}$ & 7.63 $\times$ 10$^{-1}$ & - \\
a [AU]&20234.85& 5006.19 &-\\
M$_{\rm{disc}}$ [M$_\odot$]&6.93 $\times$ 10$^5$&1.14 $\times$ 10$^2$&6.93 $\times$ 10$^5$\\
R$_{\rm{in}}$ [AU]&  1626.25&16.26&1626.25 \\
R$_{\rm{out}}$ [AU]& 4127.05&103.68&4127.05\\
\enddata
\tablecomments{Mass setup for M$_1$ and M$_2$ for each of the \scriptsize PHANTOM \normalsize simulations, mass ratio, q, and the semi-major axis of the secondary's orbit, a, for each. Note that requiring the orbital period to be 12~yrs set the radii for each simulation.  Simulation A follows the predictions made by \citet{valtonen2023}, Simulation B follows the predictions made by \citet{komossa2023}, and Simulation C is the control setup.  
}
\end{deluxetable*}

The parameters listed in Tables \ref{tab:parameters} and \ref{tab:mass_setups} describe all of the initial conditions required to setup these simulations. Because we keep the period of each simulation $\sim$12 years, the corresponding semi-major axis also changes (as seen in Table \ref{tab:mass_setups}). Extra disc parameters for a Shakura-Sunyaev disc were used such as the power law index of surface density profile $p_{index}=$3/2, the power law index of sound speed profile $q_{index}=$3/4, scale height (H/R) $=$0.05 and adiabatic gamma $\gamma=1.33$. We also used a timestep of $\sim$100 days with a full dump file output at every timestep. Simulations were ran using the the facilities of the Shared Hierarchical Academic Research Computing Network (SHARCNET) on the Rorqual cluster with 8 CPU cores.

\section{Results}\label{sect:results}

\begin{figure*}
    \includegraphics[width=\linewidth]{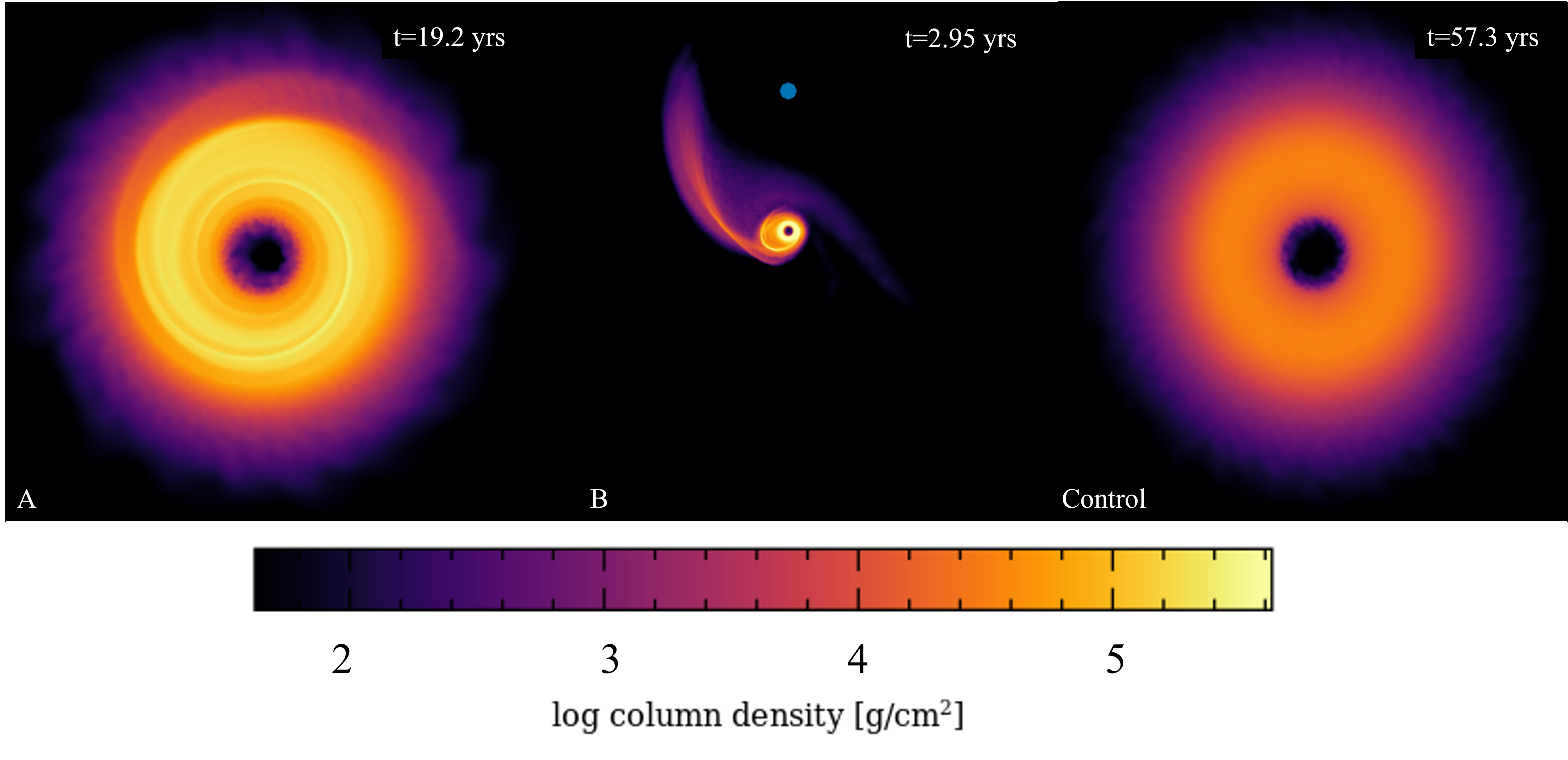}\caption{Column density snapshots of the three different simulations, in order: A, B and the Control. Simulation A shows disturbance  of the disc post impact at 19.2 years, Simulation B shows an almost fully disrupted disc at 2.95 years (the blue dot shows the location of the near-equal mass secondary SMBH) and the Control simulation shows a well-behaved, smooth disc at 57.3 years. \label{fig:allthree}}
\end{figure*}

Each of the three simulations evolved differently. Simulation A, the  currently most accepted model for the OJ287 system, showed the characteristic repeated plunging of the secondary into the disc of the primary. The control simulation went as expected with matter accreting onto the primary, single SMBH. For simulation B, the near-equal mass setup, the secondary tidally destroyed the accretion disc in under three years. Example snapshots from each simulation are shown together in Figure \ref{fig:allthree}. Simulation A is stamped at 19.2 years, simulation B at 2.95 years and the control at 57.3 years. In the control, the disc settles into a smooth, quasi-steady configuration within the first few years of the simulation without any significant variations. The snapshot at 57.3 years is representative of the duration of the simulation.

\subsection{Simulation A and Control}
\begin{figure}
    \centering
    \includegraphics[width=\linewidth]{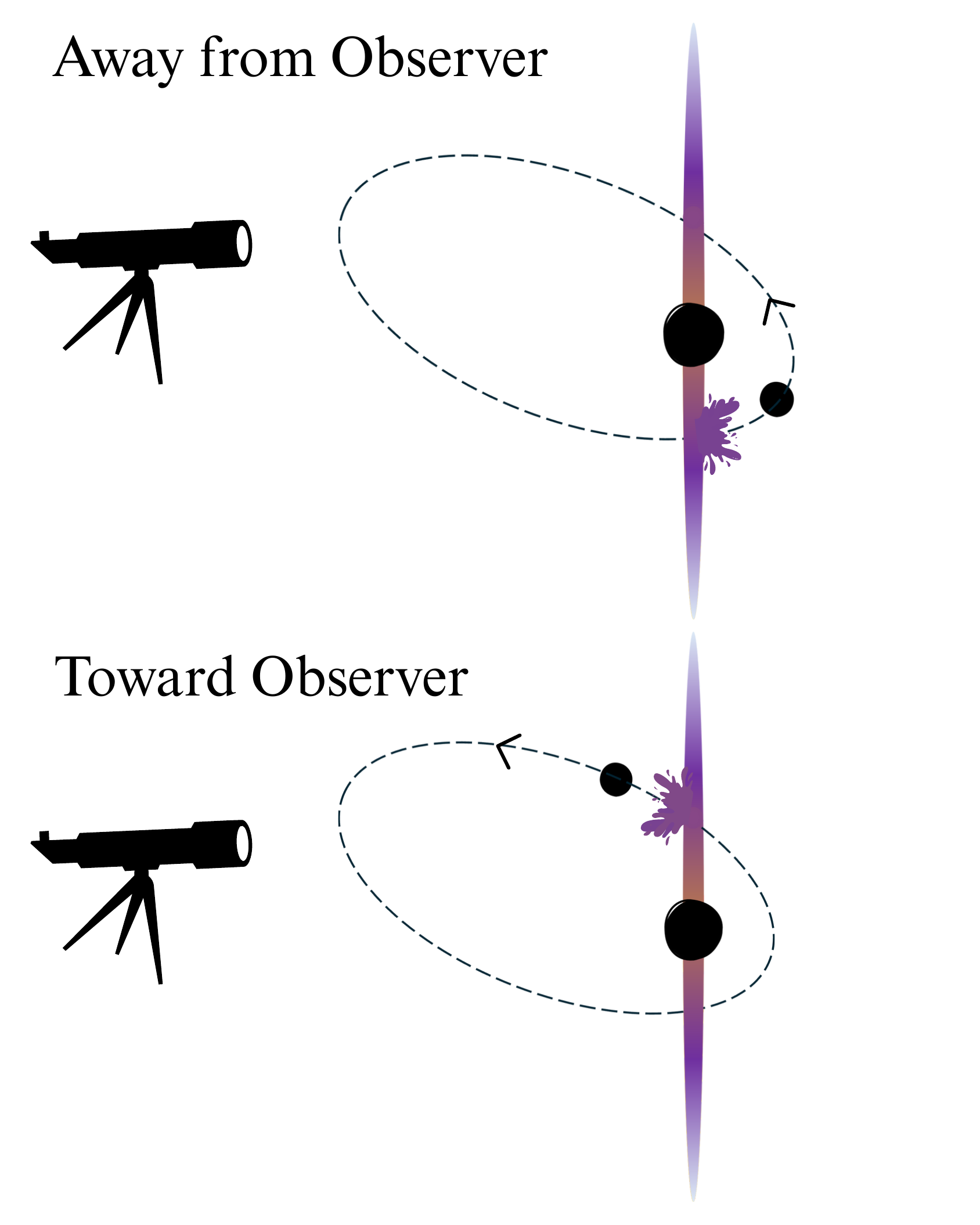}
    \caption{An illustration of the impact events of the secondary with the disc of the primary. OJ287 is a blazar; its disc is aligned parallel to the observer. During a single orbit of the secondary, one plunge pushes disc material away from us (upper diagram) and the next plunge pushes material away toward us (lower diagram).}
    \label{fig:impactdisplacement}
\end{figure}

From Simulation A, we focus primarily on the main mechanisms that could create the observed flares. The first is the interaction of the secondary with the disc that creates regions of enhanced density that temporarily boost the emissivity and accretion onto the primary.  The second mechanism is that the secondary's impact shock heats the disc and creates an expansion bubble of lower column density \citep{ivanov1998}. This, along with the gravitational pull of the secondary on the disc is responsible for the disc being disturbed vertically. An illustration of the impact of the secondary on the disc is shown in Figure \ref{fig:impactdisplacement}. 

Due to the orientation of the OJ287 system, impacts of the secondary first physically push material from the disc directly toward us and then directly away from us during a single orbit (see Figure \ref{fig:impactdisplacement}).

\subsubsection{Spiral Density Waves}
\begin{figure*}
    \centering
    \includegraphics[width=\linewidth]{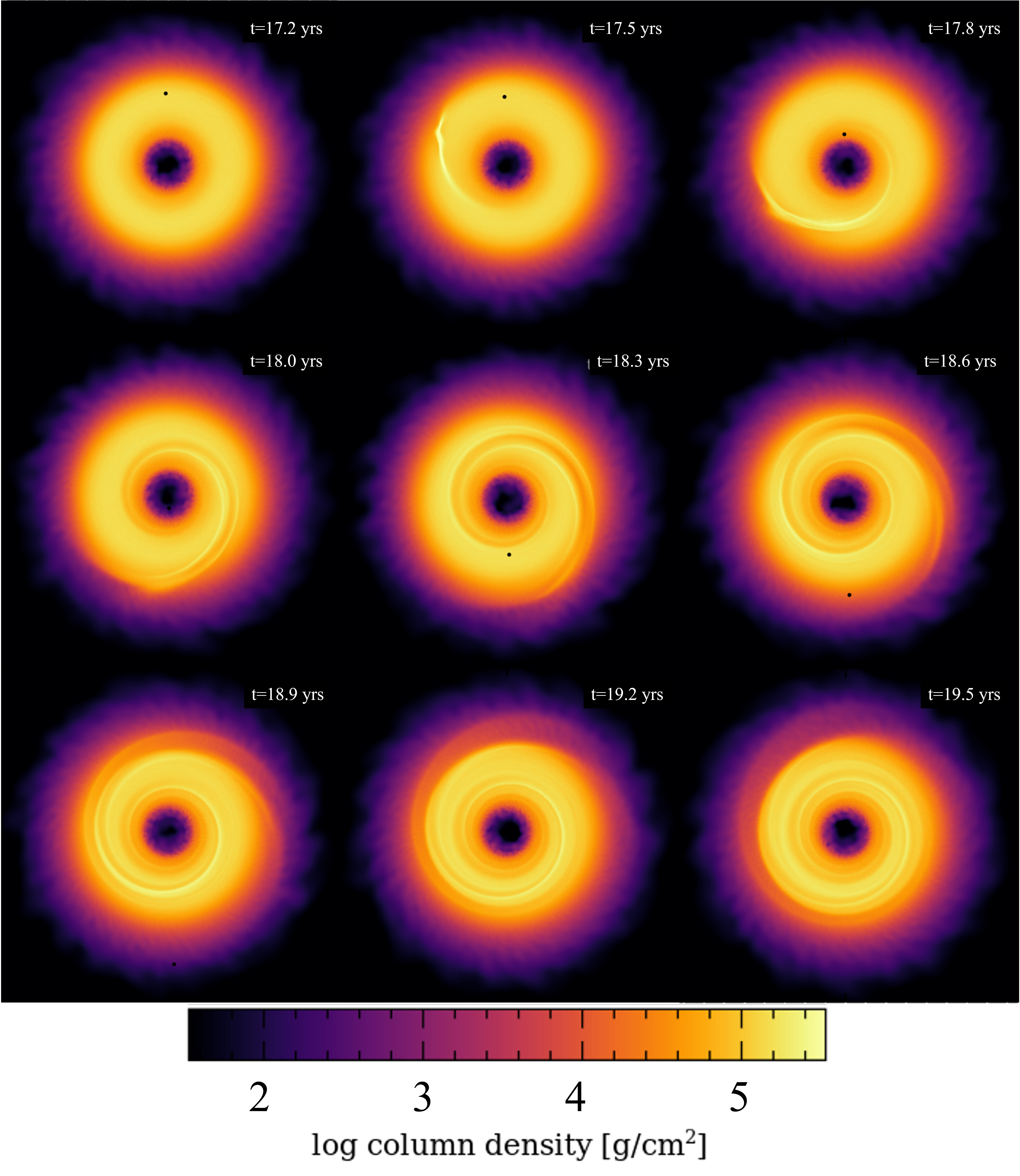}
    \caption{Column density snapshots of Simulation A just before the secondary SMBH (shown as the little black dot) of mass 1.4 $\times$ 10$^8$ M$_\odot$ hits the disc of the primary SMBH of mass 1.835 $\times$ 10$^{10}$ M$_\odot$ in the first panel at 17.2 years. At 17.5 years the secondary has hit the disc producing a point of high density. As the disc rotates a spiral density wave forms, which enables more matter to fall toward the primary SMBH in the following panels. The time from first impact to formation of the spiral is $\sim$ 0.3 years.\label{fig:simA_densityspirals}}
\end{figure*}

The most noticeable effect after the secondary's impact with the disc is the rapid formation of spiral density waves in Simulation A as shown in Figure~\ref{fig:simA_densityspirals}. The formation of spiral density waves was also observed by \citet{ressler2025} in their study of OJ287. The first impact occurs around $\sim$ 17.3 years and prominent spiral arms form within 2 years. Regions of yellow show higher column density. 
This high density forms a spiral structure that extends to the innermost disc, delivering a higher amount of material to the primary SMBH and temporarily increasing its accretion rate.

We extract the instantaneous mass accretion rate for Simulation A and normalise this to the control setup using a smoothing window of $\sim$ 200 days. Here we focus on relative variability and timing rather than reproducing the absolute OJ287 accretion rate, which depends on the disc mass normalisation, the sink (accretion) radius, and the viscosity parameter $\alpha$. Simulation A hovers slightly above the mass accretion rate for the control setup (a single SMBH in a disc) for the duration of the Simulation, but the more striking features are quasi-periodic spikes, pushing the mass accretion rate to $+$20\% the control accretion rate with a smoothing sensitivity of 200 days. These spikes are a direct consequence of the impacts and the subsequent induced spiral density waves in the disc. 

\begin{figure}
        \centering
        \includegraphics[width=\linewidth]{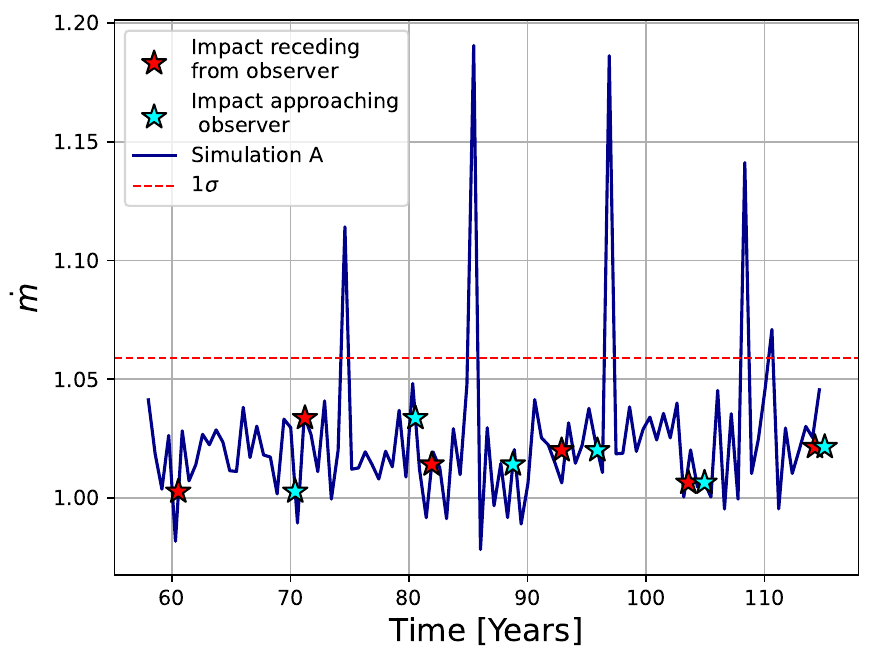}
        \caption{A zoomed in look at the accretion rate, $\dot m$, of Simulation A between 60 and 115 years. The spikes are easily seen. The red star indicates when the secondary has hit this disc heading away from us, and the blue when it has hit the disc heading toward us. A pair of these impacts appear shortly before each of these spikes though each pair of impacts do not result in a spike.}\label{fig:simAaccretionzoom}
\end{figure}

 We show a zoom-in to this accretion rate for the time-frame between 120 and 170 years of Simulation A and overlay the points at which the secondary hits the disc, in Figure~\ref{fig:simAaccretionzoom}. We see that before each spike, there is a pair of impact points. The red star indicates the time of an impact heading away from us, and the blue an impact heading toward us (see Fig.~\ref{fig:impactdisplacement}). The time lag between the second impact and the spike indicates the time taken for the spiral density wave to be accreted. The intermediary spikes in between represent the tail ends of the density spiral being accreted or just noise in the simulation. 

\begin{figure}
    \centering
    \includegraphics[width=\linewidth]{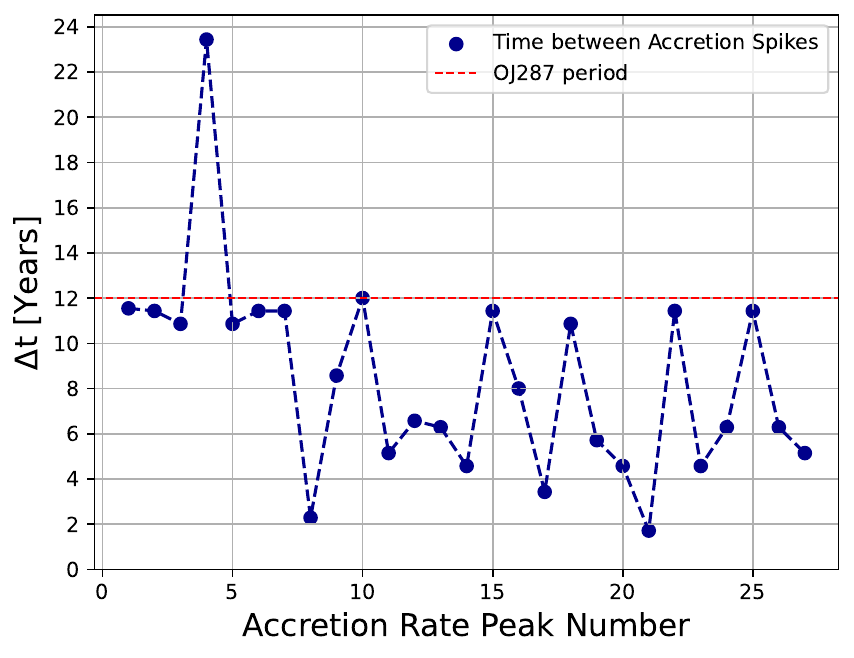}
    \caption{The time difference between successive peaks in accretion rate of Simulation A. These were selected as any peak in accretion rate above  one standard deviation from the mean accretion rate shown in Figure \ref{fig:simAaccretionzoom}. This is for the entire length of the simulation. The period of OJ287, $\sim$ 12 years, is shown as the red horizontal dashed line. There are several $\Delta t$ intervals of $\sim$12 years.}
    \label{fig:timediffaccretionpeaks}
\end{figure}

If we look at the timings between the accretion spikes, there is an average difference of $\sim$ 8 years, with standard deviation of $\sim$ 4.2 years. The times between each successive peak in the accretion rate of Simulation A selected as peaks that are greater than one standard deviation ($>$1$\sigma$) from the mean of the accretion rate normalised to the control accretion rate (as shown in Figure \ref{fig:simAaccretionzoom}) is shown in Figure \ref{fig:timediffaccretionpeaks}. The big spike seen early on, indicating a time difference of $\sim$ 24 years relates to a gap at the start of the simulation. This is because it is at the beginning of the simulation before previous impacts accumulate. The time it takes for the density waves to smooth out is typically shorter than the time between the spikes. The accretion spikes are irregular. The cumulative effect of impacts on generating spiral waves as well as precession of the secondary's orbit means that this effect is not periodic.

%

In summary, the impact of the secondary SMBH with the disc produces spiral density waves within the disc. The spiral density waves transiently enhance the accretion rate as material is pushed toward the primary. These spikes are not very large ($\sim$+20\%). Though they might contribute to the flares, from these simulations the fractional increase is too small to solely account for large flux spikes. The more likely contributor comes from the vertical disruption and from the jet, which is not simulated here. This may be a contributor to the variability of OJ287.

\subsubsection{Vertical displacement of particles}

The vertical displacement of particles has been highlighted as one of the key contributors to the optical flaring of OJ287. \citet{ivanov1998} studied the effect of non-relativistic secondary impact, both perpendicularly and inclined, on an accretion disc analytically and using 2D hydrodynamic simulations. They described how the impact of the secondary shocks the gas of the disc and as the disc material is pushed out of the disc plane it expands with a supersonic flow of material both above and below the disc. 

Our simulations are 3D and include relativistic corrections. We also explicitly simulate the OJ287 system. But we do not include radiative transfer in the models and therefore cannot directly attribute flaring to this physical, vertical displacement.

\begin{figure*}
    \includegraphics[width=\linewidth]{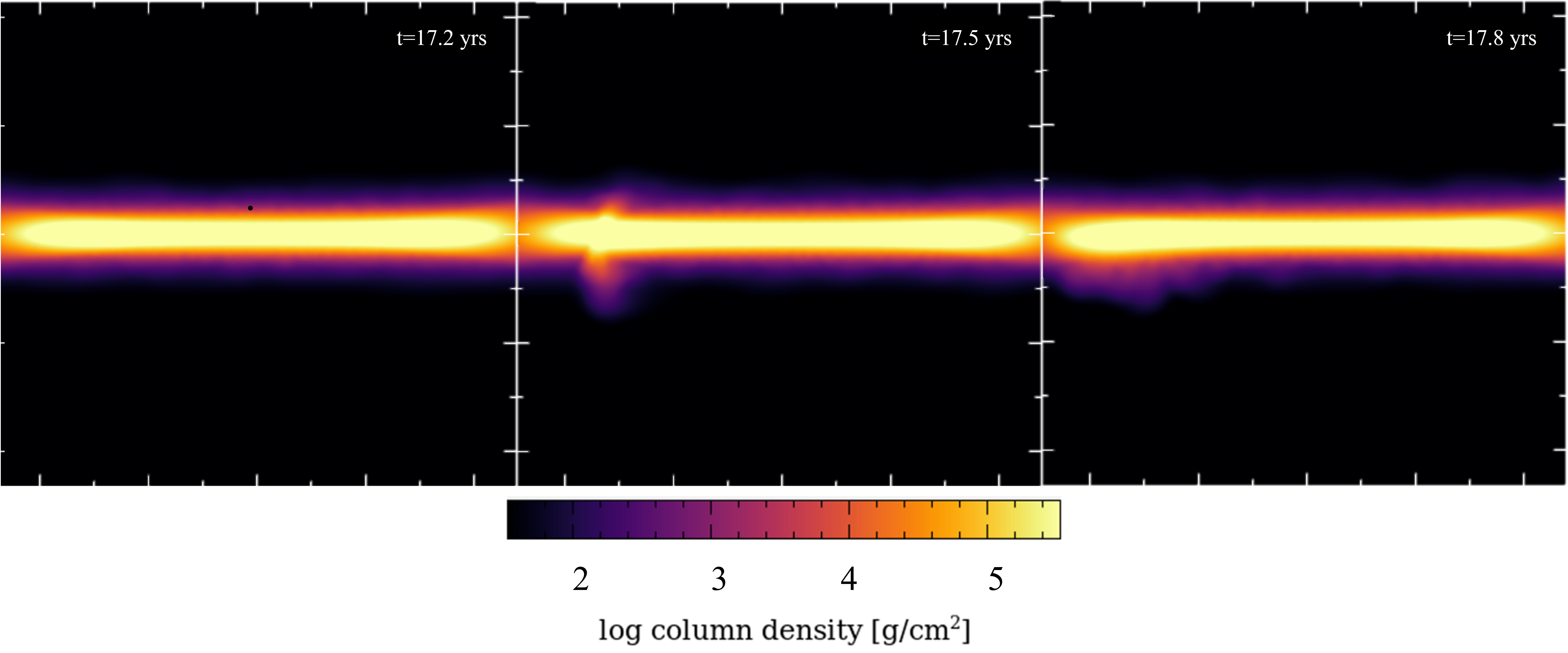}\caption{Simulation A seen from the edge-on perspective at the time of secondary disc impact. The secondary goes from the top of the disc to the bottom in these panels. The vertical ($z$-axis) disruption of particles can be seen post impact at 17.5 years (middle frame).}\label{fig:simAedgeon}
\end{figure*}

\begin{figure*}
    \includegraphics[width=\linewidth]{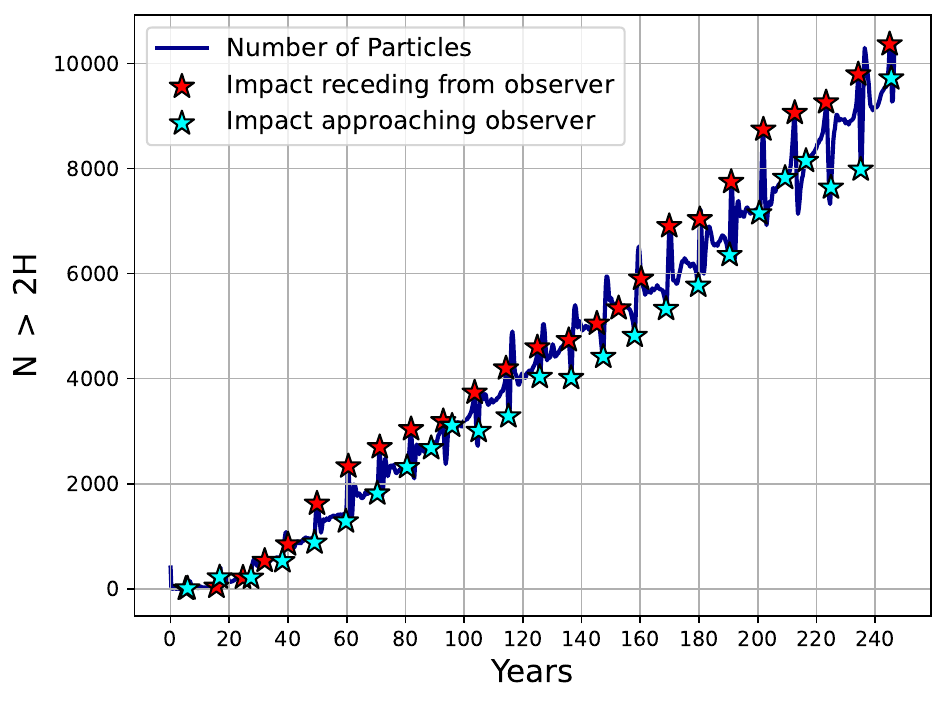}\caption{The evolution of the number of particles that go above twice the disc height of simulation A (N $>$ 2H). Stars indicate times of secondary impact. Red stars indicate impacts receding from the observer and blue stars indicate impacts approaching the observer. The underlying, smooth increase in the number of particles above twice the disc height is also seen in the control simulation.\label{fig:zcountA}}
\end{figure*}

An edge-on view of Simulation A shows the displacement of particles following the impact in Figure \ref{fig:simAedgeon}.
From Simulation A, we look at the number of disc particles that go above 2 $\times$ the disc height, $H$ (see Fig.~\ref{fig:zcountA}). Red stars highlight times of impact receding from the observer and blue stars highlight times of impact approaching the observer. It follows naturally that after an impact receding from the observer (red), the particles above disc height (facing the observer) drops and following an impact approaching the observer, the disc height rises again. While the number of particles is small, these peaks and dips can have an effect on the overall luminosity of the system. Their quasi-periodic nature shows that it is a factor that can contribute to the flaring of the system as seen in the lightcurve. 

  The increased fraction of disc particles above $2H$ primarily changes the line-of-sight column density and optical depth, which can modulate the observed flux even if $\dot{m}$ varies only modestly. This expansion of the disc particles out of the disc is though to act to increase the emissivity of the disc, lending to the flares.

\subsubsection{Orbital Effects}

\begin{figure}
    \centering
    \includegraphics[width=\linewidth]{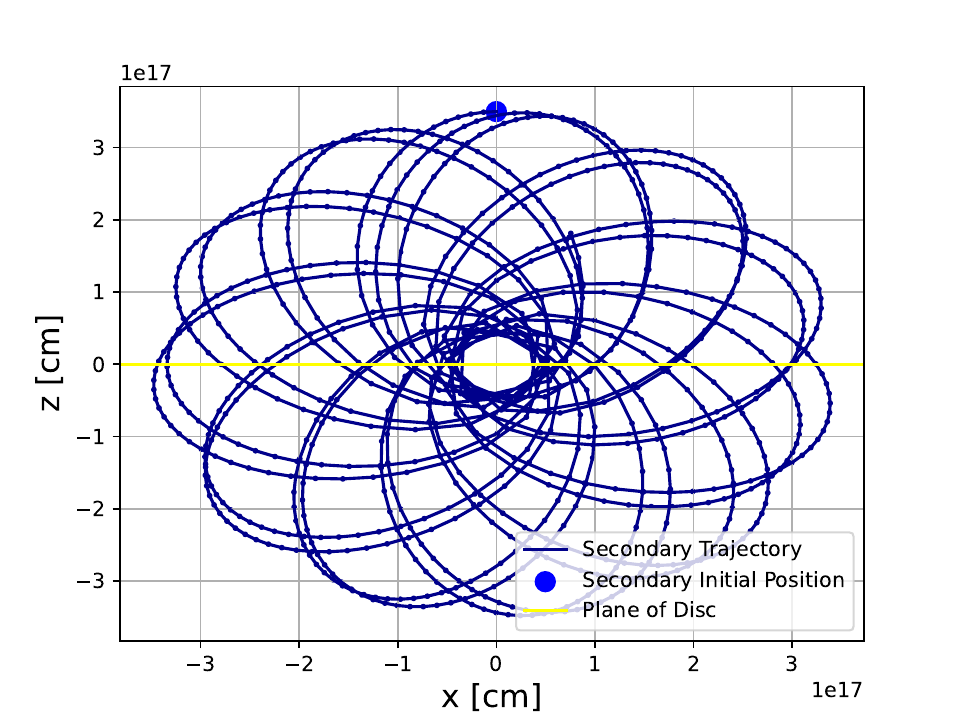}
    \caption{The orbital trajectory of the secondary SMBH in Simulation A. The horizontal yellow line represents the midplane of the disc. As the orbit of the secondary precesses the impact points along the disc occur at different radii.}
    \label{fig:secondaryprecession}
\end{figure}

A major component of the OJ287 model is the precessing nature of the system. The general relativistic precession angle, $\delta\psi$, is proportional to the total mass of the binary system \citep{einstein1915}, $M_1 +M_2$. The rate of precession of the orbit implied from the lightcurve flaring is consistent with a total mass of the OJ287 system being on the order of 10$^{10}$ M$_\odot$ \citep{lehto1996}. In our simulations, we clearly see the precession of the orbit of the secondary black hole, shown in Figure \ref{fig:secondaryprecession}.

With a precessing orbit like the one shown here, the impacts of the secondary shift in time. This affects the observed flares and is usually taken into consideration when predictions are made for observing the flares. In Table \ref{tab:impact_times}, we show the timings of the impacts from Simulation A. Here, t$_1$ (Column 2) represents the first impact of a single orbit and t$_2$ (Column 3) represents the second impact of the orbit. We find the time difference between t$_1$ and t$_2$ (Column 4) as well as the timing between successive t$_1$ impacts (Column 5). The 1 to 10 yr spread between t$_1$ and t$_2$ is a natural consequence of orbital precession and viewing geometry; this variability can explain why some expected flares appear offset in time from model predictions of a non-precessing SMBH binary. If the binary were not precessing, the timings of the impacts would all be equal with an unchanging orbit. We also include the angle of periapsis (Column 8), $\omega$, the angle between the ascending node (where the orbit crosses the orbital plane going from below to above) and periapsis (the point of closest approach to the primary).

\begin{deluxetable*}{ccccccccc}[ht!]

\tablecaption{The \textit{n} pairs of impacts during Simulation A.\label{tab:impact_times}}
\tablehead{
\colhead{n}&
\colhead{$t_{\rm 1}$} & \colhead{$t_{\rm 2}$} & \colhead{$t_{\rm 2} - t_{\rm 1}$} & \colhead{$t_{1_{i}} - t_{1_{i-1}}$} & \colhead{$t_{1_i}-t_{2_{i-1}}$}&\colhead{$\Delta t_{min}$}&
\colhead{$\omega$$^\circ$} & \colhead{R}
}
\startdata
0&5.21 & 5.75 & 0.55 & - &-&-& 267.87&1.20\\
1&15.62 & 16.71 & 1.10 & 10.41 &9.86&1.10&297.90&0.82\\
2&24.66 & 27.40 & 2.74 & 9.04 &7.95&2.74&332.80&0.68\\
3&32.05 & 38.08 & 6.03 & 7.40 &4.66&4.66&6.73&0.83\\
4&40.00 & 49.04 & 9.04 & 7.95 &1.92&1.92&40.78&0.45\\
5&49.86 & 59.73 & 9.86 & 9.86 &0.82&0.82&74.77&0.97\\
6&60.55 & 70.41 & 9.86 & 10.68 &0.82&0.82&107.41&1.40\\
7&71.23 & 80.55 & 9.32 & 10.68 &0.82&0.82&139.46&2.85\\
8&81.92 & 88.77 & 6.85 & 10.68 &1.37&1.37&167.50&5.32\\
9&92.88 & 95.89 & 3.01 & 10.96&4.11&3.01&201.83&5.02\\
10&103.56 & 104.93 & 1.37 & 10.68&7.67&1.37&220.48&3.20\\
11&114.25 & 115.07 & 0.82 & 10.68 &9.32&0.82&232.58&1.93\\
12&124.93 & 125.75 & 0.82 & 10.68 &9.86&0.82&259.43&1.39\\
13&135.62 & 136.44 & 0.82 & 10.68 &9.86&0.82&291.27&0.99\\
14&145.21 & 147.40 & 2.19 & 9.59 &8.77&2.19&325.82&0.44\\
15&152.60 & 158.08 & 5.48 & 7.40 &5.21&5.21&0.09&0.56\\
16&160.27 & 168.77 & 8.49 & 7.67 &2.19&2.19&34.01&0.89\\
17&169.86 & 179.73 & 9.86 & 9.59 &1.10&1.10&68.08&0.42\\
18&180.27 & 190.41 & 10.14 & 10.41 &0.55&0.55&101.18&0.95\\
19&190.96 & 200.55 & 9.59 & 10.68 &0.55&0.55&131.81&2.61\\
20&201.92 & 209.32 & 7.40 & 10.96 &1.37&1.37&164.88&4.94\\
21&212.60 & 216.44 & 3.84 & 10.68 &3.29&3.29&190.33&5.37\\
22&223.29 & 224.93 & 1.64 & 10.68 &6.85&1.64&201.47&3.43\\
23&234.25 & 235.07 & 0.82 & 10.96 &9.32&0.82&240.70&2.21\\
24&244.93 & 245.48 & 0.55 & 10.68 &9.86&0.55&256.18&1.29
\enddata
\tablecomments{ Times [Years] for first ($t_1$) and second ($t_2$) impacts with time differences between successive impacts ($t_2-t_1$), times between two first impacts ($t_{1_{i}} - t_{1_{i-1}}$), times between first impact and previous impact ($t_{1_i}-t_{2_{i-1}}$) and the minimum time difference between any successive impacts, $\Delta t_{min}$. Column 8 shows $\omega$, the argument of periapsis [degrees], the angle between the line of apsides and the ascending node of the orbit at the time of first impact; $R$, Column 9, is the location of the second impact in units of the initialised outer disc radius ($R_{out}$).}
\end{deluxetable*}

\begin{figure}
    \centering
    \begin{subfigure}[t]{0.45\textwidth}
        \centering
        \includegraphics[width=\linewidth]{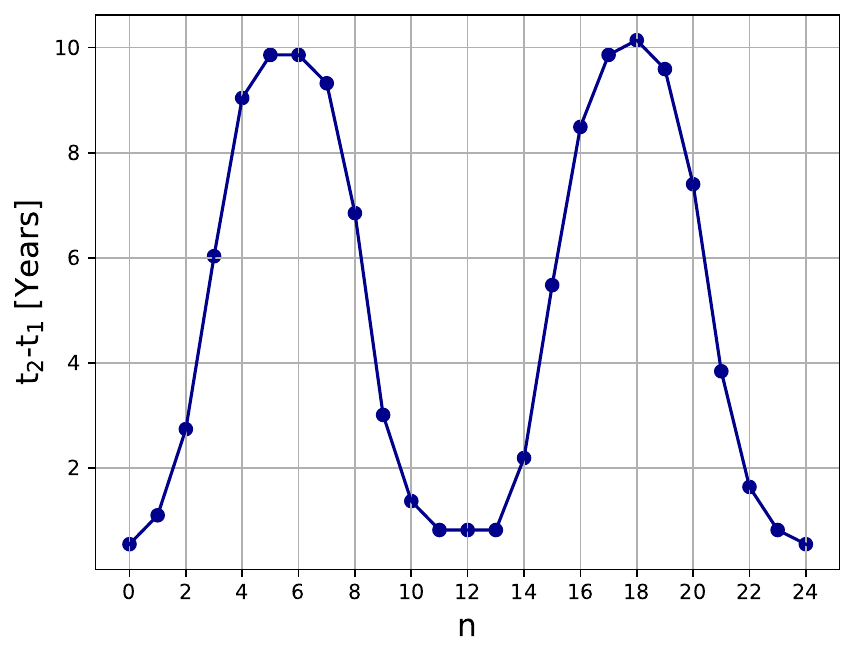}
        \caption{}\label{fig:impactdelay1and2}
    \end{subfigure}

    \vspace{1em} 

    \begin{subfigure}[t]{0.45\textwidth}
        \centering
        \includegraphics[width=\linewidth]{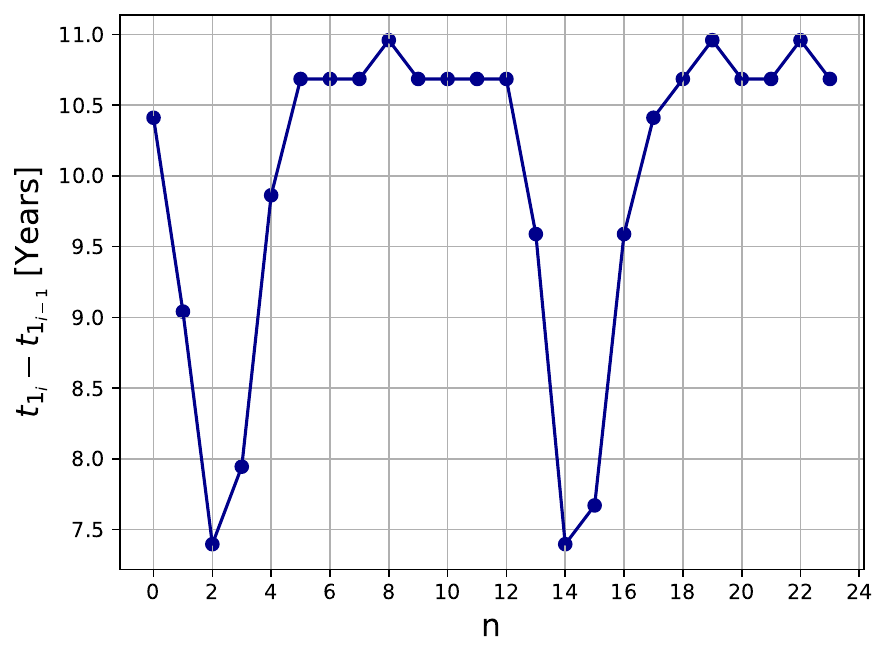}
        \caption{  }\label{fig:impactdelay1and1}
    \end{subfigure}

    \caption{The time differences between impacts of Simulation A. In the top panel we show the time difference between successive impacts $t_2$-$t_1$ (Column 4 of Table \ref{tab:impact_times}). The periodicity is a result of relativistic precession. In the bottom panel we show the time between successive first impacts of the secondary's orbit $t_{1_{i}}-t_{1_{i-1}}$ (Column 5 of Table \ref{tab:impact_times}. }
    \label{fig:two-panel}
\end{figure}

In Figure \ref{fig:impactdelay1and2}, we show the time differences between 1st and 2nd impacts of a single orbit of the secondary throughout the run of the simulation. In total there are 24 pairs of impacts. The change in time difference between the successive impacts show that the orbit is not periodic. This naturally follows the precession of the orbit of the secondary SMBH. An orbit perpendicular to the disc plane will allow the two impacts to occur close to periapsis, while an orbit lying parallel to the disc plane will allow the two impacts to be at periapsis and then at apoapsis. The timings between the two impacts vary from $\sim$ 1 year to $\sim$ 10 years (Figure \ref{fig:impactdelay1and2}). 

This large difference in the impact timings is important to consider -- especially when timing flares. The relationship between $\omega$ and impact timings is not immediately clear. Because we define the first impact of an orbit as when the secondary heads toward the disc, and away from the observer, it turns out that the longest time between first and second impact occurs when $\omega$ is $\sim$ 90$^\circ$. In this configuration the orbit is perpendicular, with periapsis above the disc. The first impact occurs with the disc, then the secondary travels along the rest of the orbit, away from the disc, then, the second impact occurs. Impact pair 6 shows this best. $\omega$ is close to 90$^\circ$ and the time between first and second impact (Column 4) is 9.86 years. But, the timing between the previous impact which was $t_2$ of n=5 and $t_1$ of n=6, is just 0.82 years (Column 6). We include Column 7 in Table \ref{tab:impact_times}, which shows the minimum time difference ($\Delta t_{min}$) between any consecutive pairs of impacts i.e. the minimum $\Delta t$ between Columns 4 and 6.

The minimum time differences, $\Delta t_{min}$, are smallest when the orbit is aligned more perpendicularly with the plane of the disc ($\omega\sim90^\circ$ and $\omega\sim270^\circ$) and largest when the orbit lies more parallel to the plane of the disc ($\omega\sim0^\circ$ and $\omega\sim360^\circ$). This aligns with the expected orbital orientations and impact timings of previous studies

We also look at the times between successive first impacts, shown in Figure \ref{fig:impactdelay1and1}. Times between two first impacts can range from $\sim$ 7 years to $\sim$ 11 years. This suggests a pattern of variability to look for. Assuming the impacts trigger the flares, then measuring the time between successful flares could offer a useful diagnostic to search for in lightcurves.

\subsubsection{Observed Flares}

Detections from OJ287 go back to the 1800's. The dates for confirmed major flares from OJ287 since 2005 are 2005.75 \citep{valtonen2008}, 2015.92 \citep{kushwaha2018} and 2019.58 (as can be seen in Figure \ref{fig:observedflux}). The time differences are 10.17 and 3.66 years, on the order of what we found for Simulation A (See Column 3, $t_2$-$t_1$ in Table \ref{tab:impact_times}).  The 2005 outburst is not in this dataset that we plot here, but we see the 2015.92 and 2019.58 outbursts. There are also some visible peaks in 2007 and 2010. The next predicted flare of OJ287 in Figure \ref{fig:observedflux} predicted by \citet{valtonen2024b} 2022, but at the time OJ287 was not visible.

\subsection{Simulation B}

\begin{figure*}
    \includegraphics[width=\linewidth]{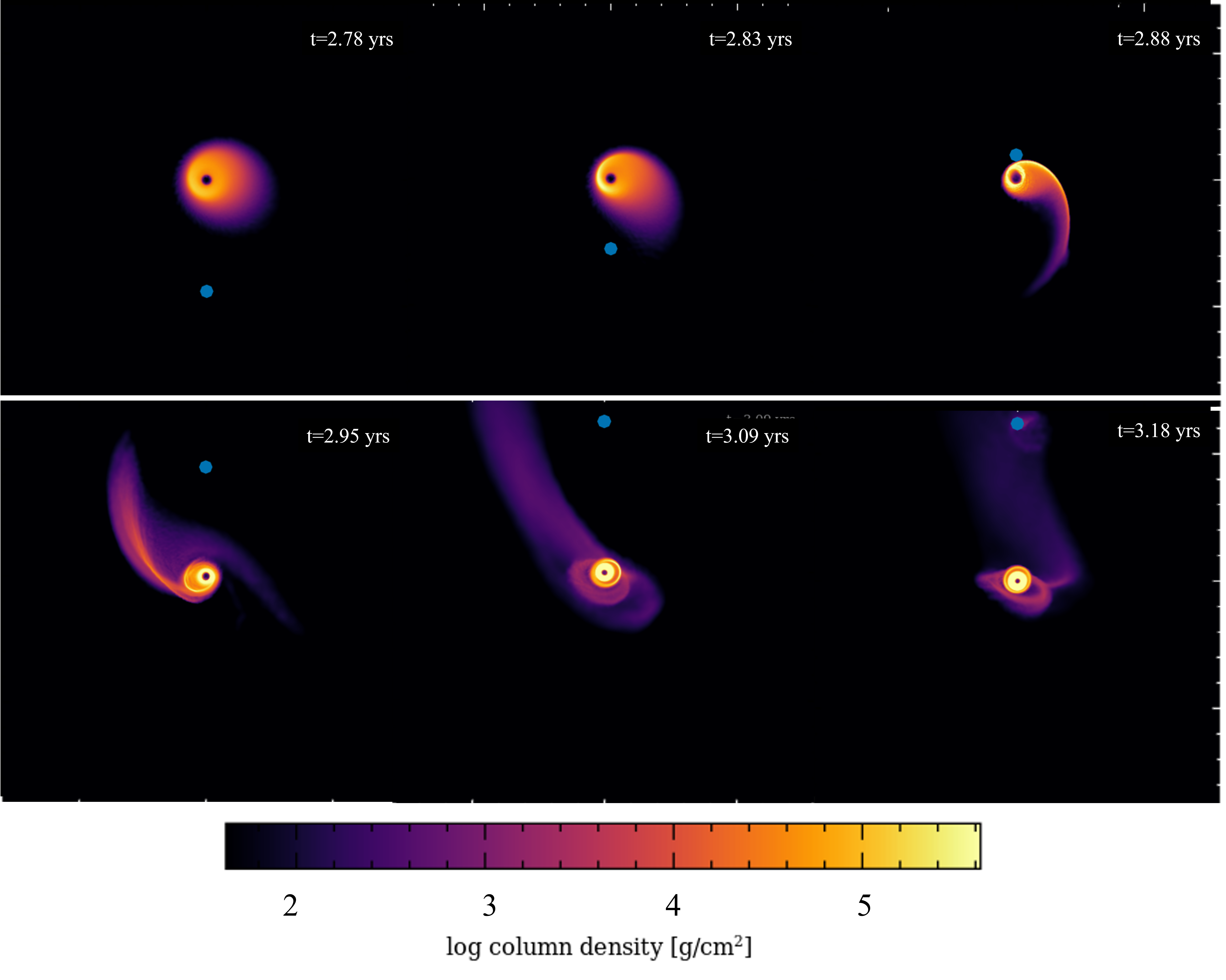}
    \caption{The evolution of Simulation B. In this simulation, the near-equal mass secondary SMBH tidally disrupts the primary's accretion disc within a few years of the start of the simulation. The secondary black hole is highlighted as a large blue dot in the panels.  }\label{fig:simBmovie} 
\end{figure*}

Simulation B evolves differently from Simulation A. The same disc geometry was used in all simulations, the main parameter changed between A and B is the mass of the primary ($\sim100$ times smaller in Simulation B).  This causes the separation to shrink for the set period of 12~yrs, and also reduces the mass of the accretion disc (see Equation \ref{eq:discmass}). From Figure \ref{fig:allthree}, we can see that the secondary tidally disrupts the disc within a few years. The time evolution of this process is shown in Figure \ref{fig:simBmovie}.  

This disruption is not surprising if we approximate the tidal disruption radius using \begin{math}
    \rm R_T=\rm R_{disc}\left[\frac{\rm M_\bullet}{\rm M_{disc}}\right]^{1/3}
\end{math}. The tidal disruption radius is $\sim 1.66 \times 10^{17}$ cm, while the initialised separation of the secondary and the primary is $\sim$5 $\times 10^{3}$AU ($\sim 7.5 \times 10^{16}$ cm; See Table \ref{tab:parameters}).  This may represent an expected evolution of equal mass SMBH binaries in which one has an accretion disc. In this scenario, the system might settle and eventually form a circumbinary disc around the two SMBHs as their orbit decays until merger \citep{armitage2002}. Based on our simulations, we find the proposed model of the near-equal mass ($\sim$10$^8$ M$_\odot$) black holes is not viable for OJ287 with a period restricted to 12 years.

\section{Discussion}\label{sect:discussion}
The two binary simulations (see Table \ref{tab:mass_setups}) evolved in dramatically distinct ways. Simulation A, which followed the \citet{lehto1996} model of the OJ287 system evolved as expected - the orbit of the secondary precessed and the disc of the primary was impacted twice per orbit. With Simulation B, the smaller, near-equal mass setup, the disc was almost immediately tidally disrupted by the secondary when we use the same orbital period as in Simulation A (see Table \ref{tab:parameters}).  It follows that Simulation B is not a viable model for OJ287.  
A significantly smaller mass primary may be possible without disrupting the accretion disc, but this would require further investigation. To our knowledge, all modeling of the secondary component of OJ287 in the literature involve a SMBH on the order of 10$^{8}$ M$_\odot$.

\subsection{Simulation A}
The timing of the impacts of the secondary SMBH with the disc of the primary in Simulation A replicates the quasi-periodicity of OJ287; the impacts also boost accretion onto the primary. However, the magnitude of this boost is small -- the accretion rate spikes briefly to $\sim$ +20\% that of the control system (see Figure \ref{fig:simAaccretionzoom}). The impacts likely contribute to the overall, long term variability of OJ287 because of the affect on the emissivity of the spiral density waves, but they do not appear to be a significant contributor to growing the primary SMBH's mass or likely to account for the observed flaring.

Spiral arm formation has been studied before by \citet{chakrabarti1993} in non-relativistic 2D hydrodynamic simulations. They simulated SMBH binaries and showed that spiral arms were induced due to the interaction with the secondary SMBH. `Hot spots' formed in the primary's disc, which they noticed could break apart and reform and lead to variability in the luminosity of the disc. 

On the observational side, \citet{storchi-bergmann2003} analysed spectroscopic observations of NGC 1097 for a period of 11 years and found that the long term evolution of the double peaked H$\alpha$ profile supports the existence of a spiral pattern in the accretion disc. NGC 1097 showed  smooth, long term variation of the double peak profile peaks in blue and red along with broadening of the profile. Broadening, they suggest, implies that the emissivity region of the disc is shifting inward, closer to the central SMBH and is therefore moving at higher velocities.

\citet{lewis2010} tried to find a universal pattern of variability in AGN discs by analysing 20 AGN line profiles. These AGN were observed for ten years with several objects being observed 2--3 times per year. \citet{lewis2010} present seven AGN profiles that exhibit variability and use two models to explain variability: i. an elliptical accretion disc and ii. spiral arm features in the disc, but found that neither model worked perfectly to describe general AGN disc variability. Though none of these objects exhibited the dramatic flaring of OJ287. 

As for the bright optical flares from OJ287, the more likely mechanism behind them is the vertical disc eruptions post impact. The emissivity of the disc is expected to increase as the gas from the disc expands outwards \citep{ivanov1998}. Confirming this explanation would require incorporating a complete radiative-transfer treatment into the simulations. 

In the work of \citet{ressler2025}, they show that the secondary also forms a short lived jet during disc impacts, though this happens only twice during the simulation. However they note that the electromagnetic contribution of the secondary's jet should be much less than that of the post impact outflows (vertical disc disruptions). 

\citet{ressler2025} also see spikes in the accretion rate post secondary impact as in our simulations. However, the magnitude of their accretion spikes (approaching 100$\times$ their control) is larger than what we observe in our simulations (1.2$\times$ our control).

\subsection{Implications for Gravitational Wave Detectors}

OJ287 is still one of the best candidates for a SMBH binary. As corroborated by our simulations, the best constraints on the masses of the black holes are $\sim10^{10}$ M$_\odot$ for the primary and $\sim10^8$ M$_\odot$ for the secondary \citep{valtonen2024b}. A binary with a primary of this size is an excellent target for pulsar time arrays (PTAs) which include  NANOGrav \citep{mclaughlin2013}, the European Pulsar Timing Array (EPTA) \citep{desvignes2016} and the Parkes Pulsar Timing Array (PPTA) \citep{manchester2013}, which form the International Pulsar Timing Array (IPTA) \citep{perera2019}. 

Though the merger timescale is $\sim10^4$ years, the short separation and high mass of this system result in a gravitational wave strain between 1.6 $\times$ 10$^{-16}$ and 9.7 $\times$ 10$^{-16}$ \citep{zhu2019, zhang2013} in the nano-hertz frequency band at present, which is within range of detection with current facilities. OJ287 would also be a contributor to the gravitational-wave background signal detected by the North American Nano-hertz Gravitational Wave Observatory (NANOGrav) measured to be $\sim$ 2.4 $\times$ 10$^{-15}$ \citep{agazie2023}. A measurement of the gravitational wave signature of OJ287 would significantly constrain the masses and orbital parameters of the system further. 



\subsection{Implications for LSST}
The OJ287 system is an interesting target in the context of the Vera C. Rubin Observatory (Legacy Survey of Space and Time (LSST)). While the orbital period of the secondary in OJ287 is $\sim$ 12 years, longer than the decade-long run of LSST, the next predicted outburst (2030's) should fall within its timeframe. It is useful to monitor the system for extended periods of time for variability outside of the impact events to both better understand the contributions from the jet as well as the more subtle affects on the light curve expected from the evolution of spiral density waves.

Detections of SMBH binaries are highly anticipated with optimistic estimates between 20-100 $\times$ 10$^6$ SMBH binaries coming from full quasar catalogues \citep{xin2021}. However, \citet{davis2024} recommend using caution when searching for SMBH binaries in the LSST quasar catalog as $\sim$ 40$\%$ of detected single quasars can be false positives for binaries due to their intrinsic variability. The notable effect of precession in this case suggests that simple sinusoidal modeling  may need to be re-evaluated when analysing highly variable lightcurves for evidence of binaries. Looking at confirmed SMBH binaries such as OJ287 with LSST will provide constraints on which characteristics of binary lightcurves could be diagnostic in their classification. This is especially so for the highly luminous quasars as simulated by \citet{davis2024}, as the more active AGN in their dataset mimicked binary signals. In their study, over 50\% of highly luminous quasars showed false positive binary signals. 

In this work, we have shown how the accretion rate varies as the impacts occur. There are spikes that arise due to the secondary's impacts with the disc. Of course, this may not be ubiquitous behaviour of SMBH binaries. In Simulation B, for near-equal mass SMBH binaries, tidal disruption of black hole discs are a distinct possibility as well. The system may evolve into a circumbinary disc (encapsulating both SMBHs) until their eventual merger \citep{franchini2024}. \citet{liao2024} report a quasar that may be such a binary embedded in a circumbinary disc by looking at the system's light curves and matching it to hydrodynamical simulations.


\section{Conclusion}
We simulated the long-term evolution of a candidate SMBH binary system, OJ287, hydrodynamically using the 3D SPH code \scriptsize PHANTOM \normalsize with two different mass ratio setups. Simulation A involved SMBHs with masses 1.835$\times$10$^{10}$ M$_\odot$ and 1.4$\times$10$^{8}$ and Simulation B involved SMBHs with near-equal mass SMBHs of 1.4$\times$10$^{8}$. The main findings from each Simulation are:

\begin{itemize}
    \item Simulation A

    (i) the impacts result in spiral density waves that mildly increase the accretion rate ($\sim$10--20\%)  $\sim$2 yrs onto the primary after paired impacts, (ii) the impacts result in splashes of material out of the mid-plane of the disc with $\Delta t$ values that are consistent with the cadence of observational flares (every $\sim$12 years)

    \item Simulation B

     the disc was tidally disrupted by the secondary around $\sim$2 years into the simulation when we use the same period as with Simulation A (see Table \ref{tab:parameters}).
    Near-equal mass black holes with this orbital period are not a viable model for this system.
    
\end{itemize}
 
The possible physical cause of the flares detected from the impact events of Simulation A were also studied in this paper. We focused on two major sources: the fraction of the disc that gets pushed up above (or down below) the surface of the disc and the impact-induced spiral density waves. From Figures \ref{fig:simAedgeon} and \ref{fig:zcountA}, the physical pushing of disc particles can be seen post impact. This shows the formation of a region of low column density and therefore low optical depth which allows for brightening of the disc for a length of $\sim$ 0.3 years. This may result in an instantaneous flaring in the disc. The spiral density waves (see Figures \ref{fig:simA_densityspirals}) are linked to the accretion rate., but are not sufficient to account for the flaring. The spikes in accretion rate lag slightly behind the impact events by $\sim$ 5 years (Figure \ref{fig:simAaccretionzoom}). \citet{ivanov1998} showed analytically and numerically that the disc itself can also become warped as the secondary approaches for impact. This can also play a role in the timing of the flares. In their work, they also described the potential formation of `fountains' of material from either side of the disc as the secondary impacts it. This is seen in Figure \ref{fig:simAedgeon}. 

We also find that the precession of OJ287 naturally describes the non-periodicity of the observational flares. The non-periodicity of impacts directly influences the non-periodicity in the temporary accretion spikes in Simulation A (See Table and \ref{tab:impact_times} and Figures \ref{fig:timediffaccretionpeaks} and \ref{fig:impactdelay1and1}).

Future work includes applying full, proper radiative transfer to model the direct impacts of the system evolution on flaring.

\begin{acknowledgments}

We would like to thank the anonymous reviewer for their quick response and helpful suggestions.
    This work was made possible by the facilities of the Shared Hierarchical Academic Research Computing Network (SHARCNET:www.sharcnet.ca) and Digital Research Alliance of Canada (https://alliancecan.ca/en).
    A.C. was supported by the Ontario Graduate Scholarship.
    A.C. and S.G. were supported by the Natural Science and Engineering Research Council (NSERC) RGPIN-2021-04157 and a Western Research Leadership Chair Award. 
\end{acknowledgments}

\section*{Author Contributions}
A.C. led the analysis and writing, S.G. and S.A. contributed to data interpretation and manuscript writing. All authors discussed and approved the manuscript for submission.

\bibliography{references}{}
\bibliographystyle{aasjournalv7}



\end{document}